\documentclass[preprint,aps,tightenlines,floatfix,showpaces,twocolumn]{revtex4}
                                                                               
\usepackage{graphicx}
\usepackage{bm}
\usepackage{amsmath}
\usepackage{amssymb}
                                                                               
\begin{document}

\title{Charge-exchange reaction cross sections and the Gamow-Teller strength 
for double beta decay}

\author{K. Amos}
\email{amos@physics.unimelb.edu.au}
\affiliation{School of Physics, University of Melbourne,
Victoria 3010, Australia}

\author{Amand Faessler}
\email{amand.faessler@uni-tuebingen.de}
\author{V. Rodin}
\email{vadim.rodin@uni-tuebingen.de}
\affiliation{Institute of Theoretical Physics, University of Tuebingen, 
72076 Tuebingen, Germany}

\date{\today}
\begin{abstract}
The proportionality between single charge-exchange reaction cross sections in 
the forward direction as found, for example  from $(p,n)$ and $(^3$He,$t)$ and 
from $(n,p)$ and $(d,^2$He) reactions, and the Gamow-Teller (GT) strength into
the same final nuclear states has been studied and/or assumed often in the 
past. Using the most physically justified theory we have at our disposal 
and for the specific example of the ${}^{76}$Ge-${}^{76}$Se system
that may undergo double beta-decay, we demonstrate that the proportionality is 
a relative good assumption for reactions changing a neutron into a proton, 
i.e. ${}^{76}$Ge$(p,n){}^{76}$As. In this channel, the main contribution to 
the GT strengths comes from the removal of a neutron from an occupied 
single-particle (SP) state and putting a proton into an unoccupied SP state 
having either the same state quantum numbers or those of the spin-orbit 
partner.  In contrast to this, in the second leg of the double beta decay a 
single proton must be taken from an occupied SP state and a neutron placed in 
an unoccupied one.  This second process often is Pauli forbidden in 
medium-heavy nuclei and only can be effected if the Fermi surface is smeared 
out.  Such is the case for ${}^{76}$Se$(n,p){}^{76}$As. Our results suggest that
one may not always assume a proportionality between 
the forward-angle cross sections of the charge-exchange reactions 
and the GT strength in any such medium-heavy nuclei.  The discrepancy 
originates from a pronounced effect of the radial dependence of the 
nucleon-nucleon ($NN$) interaction in connection with the Pauli principle 
on the cross sections in the $(n,p)$ reaction channel. Such a radial dependence is 
completely absent in the GT transition operator. 
\end{abstract}

\pacs{}

\maketitle

\section{Introduction}

In recent years, interest in the relation between charge-exchange reactions 
in the forward direction and the GT strength has increased due to the 
connection of the GT strength with the two-neutrino double beta 
decay~\cite{Ra02,Ra05}.  The latter process may help to test the nuclear 
wave functions required in calculations of the matrix elements for the 
neutrinoless double beta decay.  The neutrinoless double beta-decay 
transition probability is important since it, in conjunction with measured 
data and assuming that the light neutrino exchange is the leading 
contribution, defines an absolute scale for the mass of the Majorana neutrino 
(for reviews see, e.g.,~\cite{Fa98,El02,Mc04}).  The two-neutrino double beta
decay matrix element $M^{2\nu}$ is given by a sum over all intermediate $1^+$ states of product
of the GT transition matrix element 
from the ground state of the initial nucleus (in our example $^{76}$Ge) to 
an $1^+$ state in the intermediate nucleus ($^{76}$As) 
and the GT transition matrix elements 
from the intermediate state to the ground state of the 
final nucleus ($^{76}$Se) divided by the corresponding energy denominator.
However, a test of the two-neutrino double beta-decay calculations using the 
GT strengths extracted from the measured electron capture (EC) and the 
single-beta decay of the intermediate nucleus is only possible if the ground 
state in the intermediate nucleus is a $1^+$ state (not the case for 
$^{76}$As), and if the two-neutrino double beta decay is dominated by the 
transition through this state. Our interest with the mass-76 systems in
part stems from the fact that for the neutrinoless double beta-decay
matrix element in past evaluations gave values ranging from 2.23 to 5
according to the chosen model of nuclear structure~\cite{Ej00}. 

Complementary to the direct measurement of the GT strength by the EC capture
in the first leg followed by a $\beta^-$ transition from the lowest $1^+$ 
state of the intermediate nucleus, are single-charge transfer reactions like 
$(p,n)$ and $({}^{3}$He,$t)$ on the ground state of the initial nucleus and 
$(n,p)$ and $(d,{}^{2}$He) to the ground state of the final nucleus; all 
connecting by the intermediate $1^+$ states.  If the forward cross section of 
these charge-exchange reactions are proportional to the corresponding GT 
strength, the two-neutrino double beta-decay probability calculations can be 
checked, although the information about the relative phases of different 
contribution can not be extracted from the experimental $B(GT)$. Then it 
is possible to test the quality of the calculations for the neutrinoless 
double beta decay. 

The proportionality between the forward single charge-exchange cross section 
and the GT transition probabilities has been studied extensively in the past.  
Refs.~\cite{Ta87} and~\cite{Ra04} are particular contributions. 
Taddeucci et al.~\cite{Ta87} present an interesting analytic study of the 
proportionality involving the single charge-exchange reaction $(p,n)$ cross 
section at zero-momentum transfer which we take as quite typical of
all studies of the problem. They assumed that only angular momentum 
transfer $L=0$ is important at forward scattering angles and that the eikonal 
approximation is valid to describe the relative motion wave functions of
the incoming and emergent nucleons. Under those assumptions, they obtained an
expression for the proportionality between the forward charge-exchange 
reaction cross section and the GT transition probability, both to $1^+$ 
states.  However, in that study~\cite{Ta87} there are a number of other 
assumptions, many of which are questionable. First, a single particle-hole 
configuration is assumed for the structure of the nuclear transitions.  They 
also assume that the reaction mechanism can be taken in either a plane or a 
distorted wave impulse approximation.  Furthermore, the radial wave functions 
for the initial neutron and for the final proton in the $(p,n)$ reaction are 
assumed identical and the effects of antisymmetrization between the projectile
and the target nucleon is treated rather crudely. Limiting themselves to use 
the impulse approximation means that they use $NN$ amplitudes in calculations 
and not a specific finite ranged  $NN$ interaction.  They use expressions 
given by Franey and Love~\cite{Fr85} which were derived from the SP84 
amplitudes for free $NN$ scattering. The nuclear medium has dramatic
effects in making the effective interactions between projectile and every
bound nucleon in the target quite different to the free $NN$ case~\cite{Am00}.
Similar concerns exist even with some of the limitations are removed,
for example by using the distorted wave impulse approximation or
by using phenomenology to define relative motion wave functions in a
distorted wave approximation (DWA) approach.  Such concerns we outline 
later in the  text.

Ejiri~\cite{Ej00,Ej01} has also made extensive study of the proportionality 
link. In his review~\cite{Ej00}, the proportionality of the forward 
charge-exchange cross section for $(p,n)$ and ($^3$He,$t$) are shown in 
Figs.~10 and 15 for Fermi and GT transitions, respectively. The 
proportionality of the charge-exchange reaction cross section to the GT 
strength corresponding to a $(n,p)$ reaction is depicted in that review by 
using ($d,{}^2$He) in Fig.~12, by using ($t$,$^3$He) in Fig.~17 and by using 
($^7$Li, $^7$Be) in Fig.~18. Ejiri found that the proportionality with the 
forward scattering cross sections from single charge-exchange reactions with 
type $(p,n)$ to the GT strength was good for all nuclei to mass $A = 124$.  
However, proportionality studies for the charge-exchange reaction of the type 
$(n,p)$ was investigated only for masses to $A = 12$. Nuclei relevant for the 
double beta-decay proportionality of those charge-exchange reactions to the GT
strengths were not considered. Such are needed of course as they are important
for the two-neutrino double beta decay in the second leg where a proton 
changes into a neutron. Often that change cannot be effected by the GT 
operator $\tau^{\pm}\sigma$ which can only change particle types in orbits 
having the same quantum numbers or into the spin-orbit partner of that level.

There have been many previous studies seeking nuclear matrix elements 
from experimental data with ref.~\cite{Co06} the most recent. Cole
{\it et al.}~\cite{Co06} studied charge-exchange reactions from
${}^{58}$Ni. This nucleus is medium mass but it has a neutron excess
of only 2.  The degree of Pauli blocking to differentiate between
isospin raising and lowering transitions then is small. The case should
be classed with those of most light mass studies. The cases we consider
on the other hand have a sizeable neutron excess and so the
Pauli blocking effects in the ($n,p$) reactions are much more important
than in the ($p,n$) cases.

We consider the proportionality question again but make use of the best 
available reaction codes to evaluate cross sections for the charge-exchange 
$(p,n)$ and $(n,p)$ reactions. We consider specifically the very popular 
double beta-decay transitions $^{76}$Ge $\to$ $^{76}$As $\to$ $^{76}$Se. The 
nuclear structure of the initial, the intermediate and the final states in 
these nuclei have been defined using the Quasiparticle Random Phase 
Approximation (QRPA) with realistic forces (Bonn CD potential~\cite{Ma96}) and
with matrix elements calculated by solving the Bethe-Goldstone 
equation~\cite{Ro03,Ro06}. The results show that the forward charge-exchange 
cross section of the type $(p,n)$ for the first leg of the double beta decay 
is nicely proportional to the GT strength but that the forward reaction 
cross section of type $(n,p)$ shows rather large deviations from this 
proportionality. The latter is due to Pauli blocking since a proton from an 
occupied level must be transformed into a neutron in an empty level with the 
same ($n, \ell$) quantum numbers. Due to the radial dependence of the $NN$ 
interaction, the charge-exchange reaction can proceed by transition between SP
orbits that differ in ($n, \ell$). Such effects violate the proportionality 
between the forward charge-exchange ($(n,p)$ ($d,{}^2$He), ($t,{}^3$He), 
($^7$Li,$^7$Be) $\dots$) cross sections  and the GT strength. 

Quality of the QRPA approach for description of the GT strengths and double beta decay
has a long history~\cite{Fa98,El02,Mc04,Ro03,Ro06,Mu97,Ci05,To87,Vog86,Mu89}.
The nuclear wave functions calculated within the QRPA 
have been shown to provide good description of different properties of giant multipole resonances 
and low-lying collective $2^+$ and $3^-$ states. 
The gross structure of the GT strength distribution as well as the position of the GT resonance 
in the intermediate nuclei is correctly reproduced within the QRPA provided 
that the particle-particle strength of realistic nucleon-nucleon interaction is slightly renormalized by a factor 
$0.8\le g_{pp}\le 1.0$, depending on the model basis size~\cite{Fa98,El02,Mc04}.

The nuclear shell model, which nicely describes nuclear states in the sd shell of positive parity and 
where it is more reliable than the QRPA, is not able to describe the states relevant for double beta decay 
in the pf and sdg shells. The Strassbourg-Madrid collaboration ~\cite{Cau05} 
can only handle a basis consisting of four single-particle levels ($1f_{5/2}$,$2p_{3/2}$,$2p_{1/2}$,$1g_{9/2}$)
for the nuclei in the vicinity of $^{76}Ge$. Since the spin-orbit partners  $1f_{7/2}$ and $1g_{7/2}$ are missing 
in the model space, the model-independent Ikeda sum rule~\cite{Ikeda} is strongly violated and the GT 
strength calculations for $^{76}Ge$ within the shell model are not trustful.

The QRPA model for the nuclear wave functions is considered in the next Section
while that of charge-exchange reaction theory is developed in 
Sect.~\ref{reaction}.  Then, in Sect.~\ref{results} we present the results and
give conclusions in Sect.~\ref{conclusions}


\section{Nuclear wave functions}

The majority of calculations of the two-neutrino and the neutrinoless double 
beta decay have been made using the QRPA~\cite{Ro03,Ro06,Mu97,Ci05}. Although 
the starting points of all these studies are very similar, matrix elements 
calculated for the neutrinoless double beta-decay transition probabilities 
differ. For example, those obtained in Refs.~\cite{Ro03,Ro06,Mu97} are quite 
different from the ones of Ref.~\cite{Ci05}.  This is a reason to seek tests 
of wave functions by deriving the two-neutrino double beta-decay probability 
from the GT strengths between the initial and final nucleus to a large number 
of $1^+$ states in the intermediate nucleus. 

Herein we use wave functions obtained from QRPA calculations~\cite{Ro03,Ro06} 
in which the Brueckner reaction matrix elements of the Bonn CD potential
\cite{Ma96} for the $NN$ interaction were used.  The strength of the $NN$ 
matrix elements in the particle-particle channel has been slightly adjusted, 
by a factor $g_{pp}$, to reproduce the experimental two-neutrino double 
beta-decay probability.  For $^{76}$Ge, this value is $g_{pp} = 0.85$ for a 9 
level basis ($pf$ and $sdg$ major shells). We used the unquenched values 
$g_{ph}=1$ and $g_A = 1.25$ for the particle-hole channel renormalization 
factor and the axial coupling constant $g_A$, respectively.

Any single-particle (SP) operator of the $\beta^-$-type can be represented in second 
quantization as
\begin{eqnarray}
\label{eq1}
\beta^{-}_{JM} &=& \sum_{pn,m_p m_n} \, \langle p m_p | b_{JM} | n m_n \rangle 
\ a^{\dagger}_p a_n
\nonumber\\
&=& \hat{J}^{-1} \sum_{pn} \langle p || b_J || n \rangle 
\ C^{\dagger} (pn, JM) .
\end{eqnarray}
In this equation, $\hat{J} = \sqrt{2 J + 1}$, $C^{\dagger} (pn, JM) = \big[ 
a^{\dagger}_p\otimes \tilde{a}_n \big]^{JM}$, and $b_{JM}$ can be 
$\tau^-$ (Fermi), $\sigma \tau^-$ (GT), or any other operator, including ones 
that have $r$-dependence. The time reversed creation operator is defined as
$\tilde{a}^{\dagger}_{jm} = (-)^{j-m} a^{\dagger}_{j-m}$.  Edmond's version of
the Wigner-Eckart theorem has been used. The definition of the spherical harmonics 
includes a factor $i^l$ in order to ensure the above expression for 
the time-reversal operation. 

The reduced matrix element of such SP operators between the ground state of a 
mother nucleus and an excited state of the daughter nucleus is given by
\begin{eqnarray}
\label{eq2}
&&\langle J^{\pi} || \beta^-_J || 0^+ \rangle
\nonumber\\
&&\hspace*{1.0cm} = \hat{J}^{-1} \sum_{pn} 
\langle p || b_J || n \rangle\ \varrho^{(-)}(pn, J), 
\end{eqnarray}
where the elements of the transition matrices $\varrho^{(-)}(pn, J)$ are the
reduced matrix elements,
\begin{eqnarray}
\label{eq3}
\varrho^{(-)} (pn, J) = \langle J^{\pi}|| C^{\dagger} (pn,J) ||0^+ \rangle\ . 
\end{eqnarray}
The corresponding formulae for the $\beta^+$-channel are obtained by the 
changes
\begin{eqnarray}
C^{\dagger}(pn, J) &&\to C (pn, J)
\nonumber\\
\varrho^{(-)}(pn, J) && \to \varrho^{(+)}(pn, J) .
\end{eqnarray}

In the RPA, a nuclear state having angular momentum $J$ and projection $M$, 
is created by applying the phonon operator $Q^{\dagger}_{JM}$ to the vacuum 
state $| 0^+_{RPA} \rangle$ of the initial, even-even, nucleus, i.e.
\begin{equation}
\label{eq4}
| JM \rangle = Q^{\dagger}_{JM} | 0^+_{RPA} \rangle ;\;\;\;  
 Q_{JM} | 0^+_{RPA} \rangle = 0.
\end{equation}
Introducing the quasiparticle creation and annihilation operators, 
$\alpha^+_{\tau m_{\tau}}$ and $\alpha_{\tau m_{\tau}}$, ($\tau = p, n$) 
defined by the Bogolyubov transformation,
\begin{eqnarray}
\label{eq5}
\left( \begin{array}{c} \alpha^+_{\tau m_{\tau}} \\ 
\tilde{\alpha}_{\tau m_{\tau}} \end{array} \right)
 = \left( \begin{array}{cc} u_{\tau} & v_{\tau} \\ 
- v_{\tau} & u_{\tau} \end{array} \right) 
\left( \begin{array}{c} a^+_{\tau m_{\tau}} \\ 
\tilde{a}_{\tau m_{\tau}} \end{array} \right),
\end{eqnarray}
the phonon operator $Q^{\dagger}_{JM}$ can be written within the QRPA as
\begin{eqnarray}
\label{eq6}
Q^{\dagger}_{JM} &=& \sum_{pn} \Big[ X^{(J)}_{pn} \, A^{\dagger} (pn, JM) 
\nonumber\\
&&\hspace{1.5cm}- Y^{(J)}_{pn} \, \widetilde{A} (pn, JM) \Big],
\end{eqnarray}
where
\begin{equation}
\label{eq7}
A^{\dagger} (pn, JM) = 
\big[ \alpha^{\dagger}_p\otimes \alpha^{\dagger}_n \big]^{JM},
\end{equation}
and the forward- and backward-going free variational amplitudes $X$ and $Y$ 
satisfy the matrix equation,
\begin{equation}
\label{eq8}
\left( \begin{array}{cc} \mathcal{A} & \mathcal{B} \\ 
\mathcal{B} & \mathcal{A} \end{array} \right) 
\left( \begin{array}{c} X^m \\ Y^m \end{array} \right) = \mathcal{E}_m 
\left( \begin{array}{cc} 1 & 0 \\ 0 & -1 \end{array} \right) 
\left( \begin{array}{c} X^m \\ Y^m \end{array} \right) .
\end{equation}
Here $m$ identifies different roots of the QRPA equations for a given 
$J^{\pi}$ and 
\begin{eqnarray}
\label{eq9}
\mathcal{A} &=& \ \ 
\langle 0^+_{RPA}| \, \big[ A, \, [H, A^{\dagger}] \big]\, |0^+_{RPA} \rangle, 
\nonumber \\
\mathcal{B} &=& - 
\langle 0^+_{RPA}| \, \big[ A, \, [H, \tilde{A}] \big]\, |0^+_{RPA} \rangle.
\end{eqnarray}
For a realistic residual interaction, the matrices $\mathcal{A}$ and 
$\mathcal{B}$ are 
\begin{align}
\mathcal{A}^{J^{\pi}}_{pn, p'n'}&
= (E_p + E_n) \delta_{p p'} \delta_{n n'}
\nonumber\\
&\hspace*{0.2cm} - \Big[ g_{pp}\; G (pn, p' n'; J)
\nonumber\\
&\hspace*{1.0cm}\times
(u_p u_n u_{p'} u_{n'} + u_p u_n u_{p'} u_{n'}) 
\nonumber \\
&\hspace*{0.5cm}
 - g_{ph}\; F (pn, p' n'; J)
\nonumber\\
&\hspace*{1.0cm}\times
(u_p v_n u_{p'} v_{n'} + v_p u_n v_{p'} u_{n'}) \Big], 
\nonumber \\
\mathcal{B}^{J^{\pi}}_{pn, p' n'}& = \Big[ g_{pp}\; G (p n, p' n'; J)
\nonumber\\
&\hspace*{1.0cm}\times (u_p u_n v_{p'} v_{n'} + v_p v_n u_{p'} u_{n'}) 
\nonumber \\
&\hspace*{0.4cm}
- g_{ph}\; F (p n , p' n'; J)
\nonumber\\
&\hspace*{1.0cm}\times (u_p v_n v_{p'} u_{n'} + v_p u_n u_{p'} v_{n'}) \Big], 
\nonumber
\end{align}
where $G (p n, p' n', J)$ and $F (p n, p' n', J)$ are particle-particle and 
particle-hole interaction matrix elements of a G-matrix, respectively.

Within the QRPA, one has
\begin{eqnarray}
\label{eq10}
C^{\dagger} (p n, J M) &=& u_p v_n A^{\dagger}(p n, J M)
\nonumber\\
&&+\; v_p u_n \widetilde{A}(p n, J M) ,
\end{eqnarray}
and the transition matrix takes the form,
\begin{align}
\label{eq11}
&\varrho^{(-)} (p n, J) = 
\hat{J} \big( u_p v_n X^{(J)}_{p n} + v_p u_n Y^{(J)}_{p n} \big), 
\nonumber\\
&\varrho^{(+)} (p n, J) = 
\hat{J} \big( v_p u_n X^{(J)}_{p n} + u_p v_n Y^{(J)}_{p n} \big).
\end{align}
Correspondingly, the $B (G T)$ values for the GT transitions $0^+ \to 1^+$ can 
be written as
\begin{align}
\label{eq12}
&B (G T^{(-)}) = | \langle 1^+ || \sum_a \sigma_a \tau^-_a || 0^+ \rangle |^2
\nonumber\\
&\hspace*{0.6cm}
= \Big| \sum_{p n} \langle p || \sigma || n \rangle
\nonumber\\
&\hspace*{1.5cm} \times
(u_p v_n X^{(1^+)}_{p n} + v_p u_n Y^{(1^+)}_{p n}) \Big|^2, 
\nonumber \\
&B (G T^{(+)}) = | \langle 1^+ || \sum_a \sigma_a \tau^+_a || 0^+ \rangle |^2 
\nonumber\\
&\hspace*{0.6cm}
= \Big| \sum_{p n} \langle n || \sigma || p \rangle
\nonumber\\
&\hspace*{1.5cm}\times
(v_p u_n X^{(1^+)}_{p n} + u_p v_n Y^{(1^+)}_{p n} ) \Big|^2 .
\end{align}
In calculations, a harmonic oscillator with an oscillator length parameter 
$b = 2.09$ fm has been used to specify the SP wave functions for $^{76}$Ge and 
$^{76}$Se.  Those functions are positive at the origin.  
\begin{table}[h]
\begin{ruledtabular}
\caption{\label{QUASI} 
Two-quasiparticle configurations forming the QRPA structure of $1^+$ states in
$^{76}$As (relative to the ground state in $^{76}$Se) and corresponding 
without pairing to particle-hole states.}
\vspace{0.2cm}
\begin{tabular}{|ccc|ccc|}
&  q-p & q-h &  & q-p  & q-h  \\
\hline
ID & $n \ell j$ & $n \ell j$ & ID & $n \ell j$ & $n \ell j$ \\
\hline
1 & $0 f_{7/2}$ & $0 f_{7/2}$ & 13 & $0 g_{7/2}$ & $0 g_{9/2}$ \\
2 & $0 f_{7/2}$ & $0 f_{5/2}$ & 14 & $0 g_{7/2}$ & $0 g_{7/2}$ \\ 
3 & $0 f_{5/2}$ & $0 f_{7/2}$ & 15 & $0 g_{7/2}$ & $1 d_{5/2}$ \\
4 & $0 f_{5/2}$ & $0 f_{5/2}$ & 16 & $1 d_{5/2}$ & $0 g_{7/2}$ \\
5 & $0 f_{5/2}$ & $1 p_{3/2}$ & 17 & $1 d_{5/2}$ & $1 d_{5/2}$ \\
6 & $1 p_{3/2}$ & $0 f_{5/2}$ & 18 & $1 d_{5/2}$ & $1 d_{3/2}$ \\
7 & $1 p_{3/2}$ & $1 p_{3/2}$ & 19 & $1 d_{3/2}$ & $1 d_{5/2}$ \\
8 & $1 p_{3/2}$ & $1 p_{1/2}$ & 20 & $1 d_{3/2}$ & $1 d_{3/2}$ \\
9 & $1 p_{1/2}$ & $1 p_{3/2}$ & 21 & $1 d_{3/2}$ & $2 s_{1/2}$ \\
10 & $1 p_{1/2}$ & $1 p_{1/2}$ & 22 & $2 s_{1/2}$ & $1 d_{3/2}$ \\
11 & $0 g_{9/2}$ & $0 g_{9/2}$ & 23 & $2 s_{1/2}$ & $2 s_{1/2}$ \\
12 & $0 g_{9/2}$ & $0 g_{7/2}$ & & & \\
\end{tabular}
\end{ruledtabular}
\end{table}
Using N=3 and N=4 oscillator shells in the QRPA calculations for transitions 
to $1^+$ states in ${}^{76}$As gives a set of 23 two-quasiparticle excitations
per Eq.~(\ref{eq7}) to be included, via Eqs.~(\ref{eq6}) and (\ref{eq4}), into
the QRPA phonon creation operator for $1^+$ states in $^{76}$As
\cite{Ro03,Ro06}.  That set is shown in Table~\ref{QUASI}. The individual 
components are identified by the label ID which will be used in the discussion
of results.  Of those two-quasiparticle states, the ones labelled with ID = 5,
6, 15, 16, 21 and 22 cannot be excited by the GT operator.

\begin{center}
\begin{figure}[ht]
\scalebox{0.45}{\includegraphics*{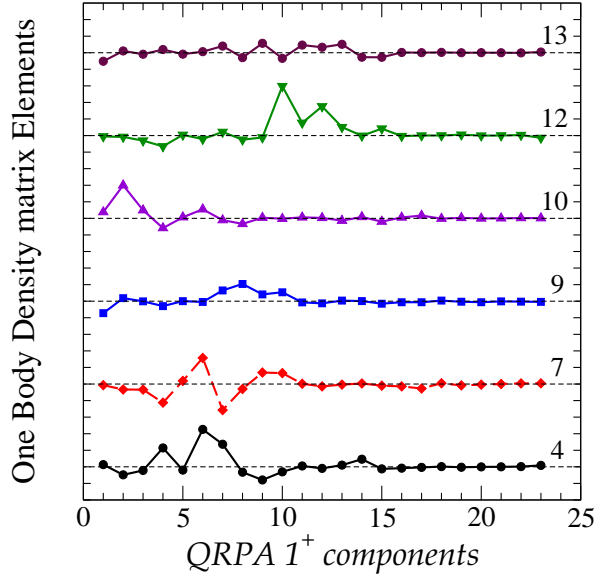}}
\caption{\label{FIG1}(Color online)
One-body density matrix elements $\rho^{(+)}(pn, 1^+)$ of Eq.~(\ref{eq11}), 
in arbitrary units, for six QRPA states 4, 7, 9, 10, 12, 13 defined in Table~\ref{200mev} 
as functions of the two-quasiparticle components 
labeled according to Table~\ref{QUASI}.}
\end{figure}
\end{center}
In Fig.~\ref{FIG1}, a set of one-body density matrix elements, $\rho^{(+)}$ 
of Eq.~(\ref{eq3})  (OBDME hereafter), for the excitation of five particular 
$1^+$ states out of the total 23 found is shown for each
of the 23 components (ID). These are the states of special interest regarding the 
${}^{76}$Se$(n,p)$ zero-degree and/or zero-momentum transfer cross sections 
considered later; being the 
strongest of the 23 charge-exchange excitations considered.  Of note is that 
the strongest OBDME $\rho^{(+)}$ of the fourth state belongs to the 
two-quasiparticle state $1p_{3/2}-0f_{5/2}$ (ID=6).  That component, readily 
excited in the charge-exchange reaction, cannot be excited by the GT operator.
Thus, one may anticipate that the proportionality between the charge-exchange 
reaction cross section and the GT strength for this state may be different to 
those of others.

\section{Reaction theory}
\label{reaction}

The charge-exchange reaction is described in a 
DWA in which one requires transition structure details, optical model wave
functions (the distorted waves), and a transition operator by which the
reaction is effected. 

\subsection{The optical potentials}
Conventionally distorted wave functions are the relative motion wave functions 
ascertained from an optical potential with which good fits to elastic 
scattering data (cross section and spin observables) have been obtained.  In 
many studies those potentials have been assumed to be local in form and usually
of Woods-Saxon type with parameters adjusted to find a good fit to elastic 
scattering data. But the associated relative motion wave functions are not 
guaranteed to be proper. Only the asymptotic (large radius) properties are 
tested by such data fitting since one only requires the scattering
phase shifts which are then used in Legendre polynomial sums that define
the cross sections etc.
There are other concerns about phenomenological model potentials to be
noted. The most serious concern is the violation of the Pauli principle.
A local potential assuming smooth parameter variations with energy will
support bound and resonance states of the compound system, and the
projectile can be captured in any. But the set includes states that are
densely occupied in the actual target. Further the resonances are then 
all single particle in nature and it is known that all nuclei support
compound and quasi-compound ones as well. For low energy scattering 
one needs a better reaction theory, such as the multi-channel algebraic
scattering theory (MCAS) to do better. Studies with that MCAS approach 
\cite{Ca05,Am05} have revealed that when coupled-channel effects are
important then the effects of violation of the Pauli principle with
potential-like models are most severe, and worse can lead to erroneous 
Physics.  This is one reason why we consider energies at which specific
coupled-channel effects, such as of virtual excitation of the giant
resonances, are minor if not negligible. Coupling to states in the
continuum~\cite{Al03} has been investigated, but, without taking into
account the effects of the Pauli principle in the optical potentials,
that result lacks some credibility. No doubt such may have a marked
role in reactions such as break-up and inelastic scattering into the
continuum, but we do not consider such coupling
to the continuum as the transitions of interest are to specific isolated states
of given spin-parity. For such, we have found no case as yet for which
the basic $g$-folding method (described next), when defined with good 
spectroscopy, requires additional reaction processes to give reasonable results.

A more physical approach is to form optical potentials by folding an $NN$ 
interaction with the target ground state structure. In that way one can also 
ensure that the Pauli principle is not violated. However, once the target 
structure has been set, then one has to choose the $NN$ interaction. For some 
time now it has been known that the interaction differs from the free $NN$ one.
Medium effects lead to the effective $NN$ interaction being energy- and 
density-dependent as well as complex.  The current best practice, at least for
energies below 3-3 resonance excitation, is to use an effective interaction 
built from the $NN$ $g$-matrices that are solutions of the 
Bethe-Brueckner-Goldstone (BBG) equations~\cite{Am00}. Using those 
$g$-matrices, both on- and off-shell values and for 32 $NN$ angular momentum 
channels, lead to an effective $NN$ interaction in coordinate space that is a 
mixture of central, two-nucleon spin-orbit, and two-nucleon tensor components.
Details of that mapping are given elsewhere~\cite{Am00}. 

Of great importance is that on using such an effective $NN$ interaction in
forming optical potentials, and when account is taken of the Pauli principle,
those optical potentials are strongly non-local and partial wave dependent. 
Non-locality arises from the allowance for knock-out (exchange) amplitudes in
the so-called $g$-folding procedure~\cite{Am00}. Doing so requires more than 
just the densities of the nuclear ground state. One requires the ground state 
OBDME
\begin{equation}
\rho_{gs}\; =\; \langle 0^+_{gs} \| 
[a_j^\dagger \otimes \tilde{a}_j]^{(J=0)} \| 0^+_{gs} \rangle .
\label{gsobdme}
\end{equation}
Assuredly, the relative motion wave functions will differ from those found
using  phenomenological (local) potentials even if the potentials are phase 
equivalent. The Perey effect is one ramification. Exchange effects are also 
most important in evaluations of non-elastic scattering and that will be 
discussed later.

In coordinate space, the $g$-folding optical potential can be written
\begin{align}
\label{amos1}
&U({\bf r}, {\bf r'}; E)
 =  \delta ({\bf r - r'}) 
\int \rho({\bf s}) \, g^D ({\bf r}, {\bf s}; E) \, \, d {\bf s} 
\nonumber\\
&\hspace*{1.6cm}
+ \sum_i n_i \, \varphi^{\ast}_i ({\bf r}) \, g^{Ex} ({\bf r}, {\bf r'}; E) \, 
\varphi_i ({\bf r'}) ;
\nonumber\\
&\;\;\;\rho({\bf s}) = \sum_i n_i\, 
\varphi^{\ast}_i({\bf s})\, \varphi_i({\bf s}) .
\end{align}
Here $\rho({\bf s})$ is the nucleon density for nucleons with the occupancies
$n_i$. To evaluate these potentials requires specification of three quantities.
They are the single nucleon bound state wave functions $\varphi_i ({\bf r})$, 
the orbit occupancies $n_i$, which more properly are the nuclear OBDME of 
Eq.~(\ref{gsobdme}), and the $NN$ $g$-matrices $g^{D/Ex}({\bf r},{\bf s};E)$. 

\subsection{The effective interaction between projectile and bound nucleons}

The $g$-matrices in the equation above, are appropriate combinations of $NN$ 
interactions in the nuclear medium for diverse $NN$ angular momentum channels. 
For those $NN$ interactions, much success has been had using an effective $NN$
interaction, now commonly designated as the Melbourne force~\cite{Am00}, and 
which has the form $g^{ST}_{01} \equiv g^{ST}_{eff} \, ({\bf r}, E; k_f (R))$ 
where ${\bf r = r_0-r_1}$ and $R=\frac{1}{2}|({\bf r_0 + r_1})|$. It is based 
on the $g$-matrix of the Bonn $B$ potential~\cite{Ma87}.  In the prescription,
the Fermi momenta relate to the local density in the nucleus at distance $R$ 
from the center when ${\bf r_i}$ are the coordinates of the colliding 
projectile and bound nucleons. $\{ST\}$ are the spin and isospin quantum
numbers of the $NN$ system.

For use in the DWBA98 program~\cite{Ra98}, these effective $NN$ $g$-matrices 
are, specifically,
\begin{align}
\label{amos2}
g^{ST}_{eff} &= g^{ST}_{eff}({\bf r}, E; k_f)
\nonumber\\
&= \sum^3_{i = 1} \Bigg[ \sum^4_{j = 1} S^{(i)}_j(E; k_f) 
\frac{e^{- \mu^{(i)}_j r}}{r} \Bigg]_{[S,T]} \, \Theta_i 
\nonumber\\
& = \sum^3_{i = 1} g^{(i) ST}_{eff}(r , E; k_f) \; \Theta_i, 
\end{align}
where $\Theta_i$ are the characteristic operators for central forces ($i = 1$),
$\{ 1, (\sigma \cdot \sigma), (\tau \cdot \tau), 
(\sigma \cdot \sigma \tau \cdot \tau) \}$, for the tensor force $(i = 2)$, 
$\{ {\bf S}_{12} \}$, and for the two-body spin-orbit force $(i = 3)$, 
$\{ {\bf L} \cdot {\bf S} \}$. The $S^{(i)}_j(E;k_f)$ are complex, energy- and
density-dependent strengths. The properties of the $g$-matrices are such that,
not only can the ranges of the Yukawa form factors be taken as independent of 
energy and density~\cite{Am00}, but also four suffice with this approach for 
energies to just below the 3-3 resonance threshold. 

The strengths (and ranges) in these effective $NN$ interactions were found by 
mapping their double Bessel transforms to the $NN$ $g$-matrices in infinite 
nuclear matter (solutions of the BBG equations). 
With $\alpha$: $\{L L' J S T \}$, this mapping is
\begin{equation}
\label{amos3}
g^{J S T}_{\mbox{eff}; L L'}(q', q; E) 
= \sum_i \langle \Theta_i \rangle\; {\cal I}_i ,
\end{equation}
where the radial integrals expand to
\begin{align}
{\cal I}_i &=
\int^{\infty}_0 r^{2 + \lambda} j_L(q' r)\; 
g^{(i) S T}_{\mbox{eff}}(r, E; k_f)\; j_{L'}(q r) \, d r 
\nonumber \\
&= \sum_{j} S^{(i)}_j (\omega)
\nonumber\\
&\;\;\;\;\;\;\times
 \int^{\infty}_0 r^{2 + \lambda} 
j_L(q' r) \frac{e^{- \mu^{(i)}_j r}}{r} j_{L'}(q r) d r \nonumber \\
&= \sum_{j}  S^{(i)}_j(\omega)\; 
\tau^{\alpha}(q', q; \mu^{(i)}_j).
\end{align}
Therein $\lambda = 2$ for the tensor force. In application, a singular valued 
decomposition has been used to effect this mapping.

\subsection{The DWA for non-elastic reaction analyses}

In the DWA, amplitudes for a non-elastic scattering of nucleons from
nuclei, through a scattering angle 
of $\theta$, and between the states $\big| J_i, M_i \big\rangle$ and 
$\big| J_f, M_f \big\rangle$, are
\begin{align}
\label{amos4}
&T_{DWA} = T^{M_f M_i \nu' \nu}_{J_f J_i}(\theta) 
\nonumber\\
&\;\;= \bigg\langle \chi^{(-)}_{\nu'}({\bf k}_0 0) \Big| \Big\langle 
\Psi_{J_f M_f}(1 \cdots A) \Big| 
\nonumber \\
&\;\;\;\; \times A \sum_{S T} g_{eff}^{S T}({\bf r}_{0,1}, E; k_f) \, P_S P_T 
\nonumber\\
&\;\;\;\;\;\;\;\;\;\times
\mathcal{A}_{01} \, \{ \Big| \chi^{(+)}_{\nu}({\bf k}_i 0) \Big\rangle \Big| 
\Psi_{J_i M_i}(1 \cdots A) \Big\rangle \},
\end{align}
where $\nu$,$\nu'$ are the spin quantum number of the nucleon in the continuum,
$\chi^{(\pm)}$ are the distorted waves, and $g^{S T}_{eff}({\bf r}_{0,1},E;k_f 
\, P_S P_T$ is the spin-isospin Melbourne force.  The operator 
$\mathcal{A}_{01}$ effects the antisymmetrization of the two-nucleon product 
states.

Then, by using cofactor expansions, $|\Psi_{J M} \rangle = A^{-1/2} 
\sum_{j, m} | \varphi_{j m}(1)\, \rangle \, a_{j m} | \Psi_{J M} \rangle$, 
the matrix elements become
\begin{align}
\label{amos5}
&T^{M_f M_i \nu' \nu}_{J_f J_i} = \sum_{j_1, j_2 i, S, T} 
\langle \Psi_{J_f M_f}|a^{\dagger}_{j_2 m_2} a_{j_1 m_1}|\Psi_{J_i M_i}\rangle 
\nonumber \\
&\;\;\ \times \Big\langle \chi^{(-)}_{\nu'}({\bf k}_0 0) \Big| \Big\langle 
\varphi_{j_2 m_2}(1)\Big| \, 
g^{S T}_{eff}({\bf r}_{0,1}, E; k_f)
\nonumber\\
&\;\;\;\;\;\;\times P_S P_T \mathcal{A}_{01} \, 
\Big\{ \Big| \chi^{(+)}_{\nu}({\bf k}_i 0) \Big\rangle \, \Big| 
\varphi_{j_1 m_1}(1) \Big\rangle \Big\} .
\end{align}
The density matrix elements in the amplitudes reduce as
\begin{align}
\label{amos6}
&\bigg\langle \Psi_{J_f M_f} \bigg| a^{\dagger}_{j_2 m_2} \, a_{j_1 m_1} 
\bigg| \Psi_{J_i M_i} \bigg\rangle 
\nonumber \\
&\;\;\;=\sum_{I (N)} (- 1)^{(j_1 - m_1)} \, 
\langle j_1, j_2, m_1, - m_2 | I, N \rangle 
\nonumber\\
&\;\;\;\;\;\;\times 
\Big\langle \Psi_{J_f M_f} \Big| \bigg[ a^{\dagger}_{j_2} \otimes a_{j_1} 
\bigg]^{I N} \Big| \Psi_{J_i M_i} \Big\rangle 
\nonumber \\
&\;\;\;=\sum_{I (N)} (- 1)^{(j_1 - m_1)} \, 
\langle j_1, j_2, m_1, - m_2 | I, N \rangle 
\nonumber\\
&\;\;\;\;\;\;\times 
\langle J_i, I, M_i, N | J_f, M_f \rangle \, \frac{1}{\sqrt{2 J_f + 1}} 
\, S_{j_1 j_2 I} , 
\end{align}
where $S_{j_1 j_2 I}$ are the transition OBDME. The DWA amplitudes are then
\begin{align}
\label{amos7}
&T^{M_f M_i \nu' \nu}_{J_f J_i} = \sum_{\xi} 
\frac{(-)^{(j_1 - m_1)}}{\sqrt{2 J_f + 1}} \, S_{j_1, j_2, I} 
\nonumber\\
&\times
\langle j_1, j_2, m_1, - m_2 | I, N \rangle 
\, \langle J_i, I, M_i, N | J_f, M_f \rangle 
\nonumber \\
&\;\;\;
\times \Big\langle \chi^{(-)}_{\nu'}({\bf k}_0 0) \Big| 
\, \langle \varphi_{j_2 m_2}(1) | {\bf g}^{S T}_{eff}({\bf r}_{0,1}, E; k_f) 
\nonumber\\
&\;\;\;\;
\times P_S P_T \mathcal{A}_{01}\, \bigg\{ \bigg| \chi^{(+)}_{\nu}({\bf k}_i 0) 
\bigg\rangle \, | \varphi_{j_1 m_1}(1) \rangle \bigg\}. 
\end{align}
In this, $\{ \xi \} = j_1, j_2, m_1, m_2, I (N), S, T$ with $j_2$ being the 
particle and $j_1$ the hole in a particle-hole specification of the transition.

Thus, in our DWA  evaluations of the charge-exchange scattering of interest, 
namely $^{76}$Ge$(p,n)$ and $^{76}$Se$(n,p)$ to $1^+$ states given by a QRPA 
model, we have used
\begin{enumerate}
\item
SP wave functions: harmonic oscillators with oscillator length of 2.09 fm. 
Those are used to specify both the optical potentials and the reaction 
amplitudes.
\item
Optical potentials (to give the distorted waves) are formed with the Melbourne 
effective $NN$ interaction at the relevant incident particle energies. The 
occupancies of the single particle level are automatically given in the QRPA 
due to pairing and configuration mixing and are contained in the OBDME in 
Eq.~(\ref{eq11}).
\item
The same effective interactions are used in evaluations of the charge-exchange
cross sections.
\item
The $\rho^{(\pm)}$ of Eq.~(\ref{eq11}) are taken as the $S_{j_1,j_2,I = 1}$ 
depending upon which reaction, $(p,n)$ or $(n,p)$, is described.
\end{enumerate}


\section{Results}
\label{results}

We have stressed the importance of using an appropriate optical model to 
define the distorted wave functions in DWA evaluations of the charge-exchange 
scattering. We contend that potentials formed using the $g$-folding procedure 
are such for incident nucleon energies in the range $\sim 40$ to $\sim 300$ 
MeV and for nuclei for which, at the minimum, sensible models of their ground 
state structures can be specified.

\begin{center}
\begin{figure}[ht]
\scalebox{0.45}{\includegraphics*{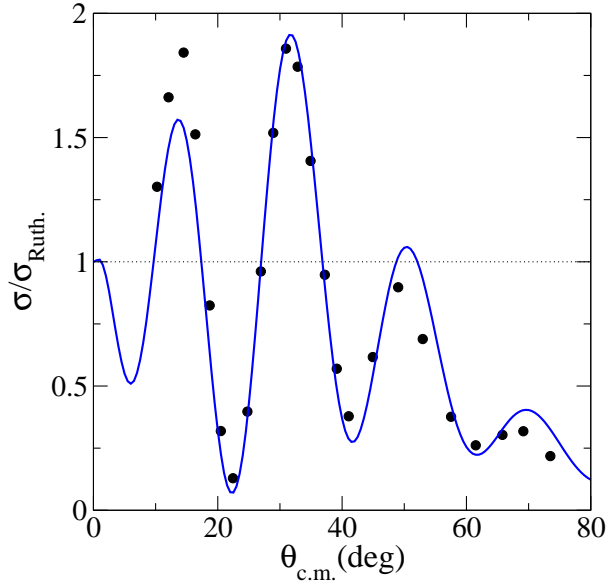}}
\caption{\label{FIG1.5}
Ratio to Rutherford cross sections for 64.5 MeV protons scattering from 
${}^{76}$Se. The solid line represents the calculation result of this work.}
\end{figure}
\end{center}

For the mass-76 nuclei we consider specifically, very few nucleon scattering 
results have been reported.  We have found data for the scattering of 22.3 MeV
protons from both ${}^{76}$Ge and ${}^{76}$Se~\cite{Mo93} and for 64.5 MeV
proton scattering from ${}^{76}$Se~\cite{Og86} (see fig.~\ref{FIG1.5}). For nucleon scattering off of
these targets, 22.3 MeV may be too low an energy to have confidence that the
reaction processes not included in the $g$-folding method, e.g. coupled-channel
effects, multi-step processes, and the like, may have importance. For energies 
40 MeV and higher, such extra processes have not been needed to find good 
replication of elastic scattering data with  $g$-folding model
evaluations~\cite{Am00,La01,De01,St02,De05,Am05a,Ki06}, when good structure, 
and appropriate effective $NN$ interactions are used,  and a proper treatment 
of the Pauli principle is made. 

In this study, we have chosen four incident energies at which the transition
to the lowest $1^+$ state in ${}^{76}$As is considered. Those energies are 45,
65, 120, and 200 MeV.  Subsequently, for our investigation of transitions to 
all 23 $1^+$ state excitations, we have used just the two largest energies of 
120 and 200 MeV.  Proton elastic scattering data have been taken taken for all
four energies and from many targets. With most cases, $g$-folding model analyses
\cite{Am00} gave good reproductions of the observations, especially whenever 
good models for the structure of the target were available.  Besides results 
given in the review~\cite{Am00}, in more recent studies the $g$-folding method 
has been used to 
assess the neutron excess distributions in nuclei~\cite{Ka02,Kl03,Am04,Am06},
to compare with non-relativistic and relativistic phenomenological
Schr\"odinger equation solutions~\cite{De01,De05}, and to ascertain
neutron halo or neutron skin characteristics in light mass radioactive 
nuclei~\cite{La01,St02}.

It is a mantra of $g$-folding studies that no adjustments to details specified 
are considered {\it post facto}. Consequently, $g$-folding predictions 
invariably do not yield the quality of fit to a data set that may be obtained 
by appropriate adjustment of parameters in current, phenomenological, optical 
potentials~\cite{Ko03}.  Nonetheless, the $g$-folding approach does give cross
sections that compare well with observations; well enough that, in some cases, 
results~\cite{La01,St02,Ka02,Am04,Am06} revealed whether a nucleus had a 
neutron skin or halo.
An analysis of cross sections from the elastic scattering of 65 MeV protons
from ${}^{76}$Se is illustrative of the quality of the $g$-folding potential 
results.  That data~\cite{Og86}, in ratio to Rutherford form, are compared 
in Fig.~\ref{FIG1.5} with the (single calculation) result from our $g$-folding 
optical model potential of the system.  With the exception of the forward peak,
our prediction compares favorably with the result of the phenomenological 
optical potential calculation of Ogino {\it et al}~\cite{Og86}. The agreement 
with data from our non-phenomenological approach suffices to give confidence 
that the non-local, complex, optical potential formed fully microscopically is 
a credible, physically justified, one. It is important to note that there is no
addition of any phenomenological elements as used in what may be termed 
semi-microscopic methods~\cite{Al03,Jo05}. 

\subsection{$(p,n)$ reactions to $1^+$ states in ${}^{76}$As}

\begin{center}
\begin{figure}[ht]
\scalebox{0.4}{\includegraphics*{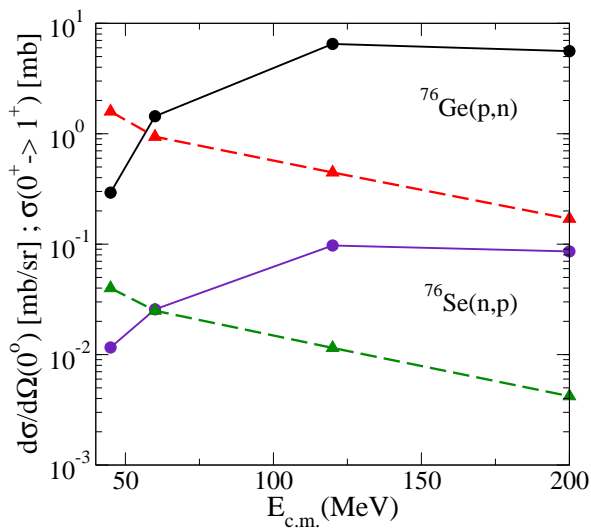}}
\caption{\label{FIG2}(Color online)
Zero-degree differential cross sections (filled circles) and total reaction 
cross sections (filled triangles) from $^{76}$Ge$(p,n)$ and $^{76}$Se$(n,p)$ 
leading to $^{76}$As($1^+_1$, 0.947 MeV).} 
\end{figure}
\end{center}

\begin{center}
\begin{figure}[t]
\scalebox{0.4}{\includegraphics*{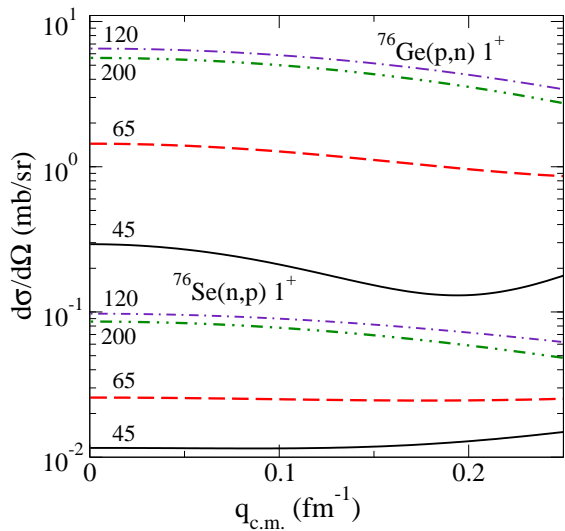}}
\caption{\label{FIG3}(Color online)
Differential cross sections from the two charge-exchange reactions leading to
the first $1^+$ state in ${}^{76}$As for bombarding energies of of 45 (solid 
curves), 65 (dashed curves), 120 (dot-dashed curves) and 200 (double dot-dashed 
curves) MeV.}
\end{figure}
\end{center}

Differential cross sections  evaluated at zero-degree scattering and the total 
reaction cross sections from ${}^{76}$Ge$(p,n)$ and ${}^{76}$Se$(n,p)$ leading
to the first $1^+$ state in ${}^{76}$As are displayed in Fig.~\ref{FIG2}. 
The results found at energies of 45, 65, 120, and 200 MeV, are connected by 
solid lines ($0^\circ$ cross sections) and by dashed lines (reaction cross 
sections).  The results of $^{76}$Ge$(p,n)$ to the first excited $1^+$ state 
in $^{76}$As are larger than those of $^{76}$Se$(n,p)$ to the same first 
excited $1^+$ state.  Over these energies, those ratios range from 25 to 65. 
It is intriguing that both the zero-degree differential cross sections, which 
increase with energy, and the reaction cross sections, which decrease 
accordingly, have such similar ratios.  
The scale factors are not simply a zero-degree phenomenon. This is emphasized 
by the differential cross-section results for the four energies displayed in 
Fig.~\ref{FIG3} for a small range of momentum transfer $q_{cm}$ from 0.  The 
results in the top of this 
figure for the four energies as indicated, are those from the charge-exchange 
$(p,n)$ reaction. The other (smaller in magnitude) results are differential 
cross sections for the $(n,p)$ reaction.  The scale factors are an effect of 
the Pauli principle.  In $^{76}_{32}$Ge$_{44}$, the proton Fermi surface lies 
between the $1p_{3/2}$ and the $0f_{5/2}$ levels while the neutron Fermi 
surface lies within the $0g_{9/2}$ single-particle state. For the $(p,n)$ 
charge-exchange reaction, one must move a neutron into a proton level. In 
these nuclei the GT transition operator $\tau^-\sigma$ can make a nucleon into
a single particle level with the same quantum numbers or to the spin-orbit 
partner. This is possible for transitions $0f_{5/2} \to 0f_{5/2}$, $1p_{1/2} 
\to 1p_{1/2}$ and $0g_{9/2} \to 0g_{9/2}$. But for the inverse
reaction, $(n,p)$ on $^{76}_{34}$Se$_{42}$, all the possible single-particle 
GT transitions are strongly Pauli hindered, if not Pauli blocked.  The latter 
cases allow GT transitions since the GT operator can only move a proton into a
corresponding neutron level of the same ($n \ell j$) orbit or the spin-orbit 
partner because of  the smearing of the Fermi surface which is mainly induced 
by pairing correlations. For the charge-exchange reaction $(n,p)$ on the other
hand, the finite range character of the transition operator and the knock-out 
process associated with antisymmetrization allow non-GT type transitions to 
contribute.  
The energy variation in magnitudes of these results for each reaction
separately reflects the energy dependence of the contributing terms to the
charge exchange process in the effective $NN$ interaction. Tensorial
components, which do not contribute strongly to the charge exchange
processes we consider herein, nonetheless vary in importance in the overall 
prescription of the medium modified force~\cite{Am00} with energy and with 
angular momentum transfer~\cite{Am05a,Ki06}.  Above 100 MeV these cross 
sections are essentially of the same magnitude and so we consider the two 
energies, 120 and 200 MeV, in the subsequent discussions.  We note that, for 
all four energies, and for those 
of 120 and 200 in particular, cross sections smoothly decrease away from
the zero-momentum transfer values over the small momentum transfer values
considered. Furthermore, while the momentum transfer values for zero-degree 
scattering at these two energies are small but non-zero, evaluations of cross 
sections setting energies to give a zero-momentum transfer gave values less 
than a percent different from those of the actual zero-degree calculations.

The foregoing dealt only with the excitation of the first $1^+$ state in
${}^{76}$As. We now consider the zero-degree cross sections for all 23 possible
$1^+$ states defined by the QRPA and their ratios to the corresponding
GT strengths. Shown in table~\ref{200mev} those cross-section values for the 
$^{76}$Ge$(p,n)$ and $^{76}$Se$(n,p)$ reactions to each of the 23 excited 
$1^+$ states are listed in columns 2 and 4. The ratios of those with the 
corresponding GT strengths are listed in columns 3 and 5.
\begin{center}
\begin{table}[t]
\begin{ruledtabular}
\caption{\label{200mev} differential cross sections at zero degrees scattering
for the charge-exchange reactions $^{76}$Ge$(p,n)$ and $^{76}$Se$(n,p)$
exciting $^{76}$As$(1^+, m)$.  The projectile energy in all cases was 200MeV. 
${\cal r}$ are the ratios of each of those cross sections with the associated,
dimensionless, GT strength. The calculated excitation energy $E_x$ of the $1^+$
states is measured from the ground state of $^{76}$Ge.}
\vspace{0.2cm}
\begin{tabular}{|c|c|cc|cc|}
$m$ & $E_x$, MeV &$^{76}$Ge$(p,n)(0^{\circ})$ & ${\cal r}$ & $^{76}$Se$(n,p)(0^{\circ})$ & ${\cal r}$ \\
\hline
1 & 1.16 & 5.61 & 3.73 & 0.09 & 3.15 \\
2 & 2.10 & 1.74 & 5.55 & $1 \times 10^{-3}$ & 25.48 \\
3 & 2.41 & 1.01 & 5.27 & 0.03 & 10.96 \\
4 & 2.87 & 2.35 & 6.19 & 0.22 & 6.93 \\
5 & 3.14 & 1.07 & 3.80 & $3 \times 10^{-4}$ & 571.43 \\
6 & 3.41 & 3.04 & 4.95 & $3 \times 10^{-3}$ & 40.12 \\
7 & 3.95 & 12.28 & 4.16 & 0.26 & 3.71 \\
8 & 4.86 & 2.31 & 3.56 & 0.08 & 3.54 \\
9 & 5.16 & 15.01 & 3.96 & 0.38 & 3.10 \\
10 & 6.39 & 0.34 & 5.45 & 0.18 & 6.27 \\
11 & 8.40 & 18.43 & 3.47 & 0.04 & 3.40 \\
12 & 9.90 & 4.77 & 4.09 & 0.31 & 3.83 \\
13 &11.25 & 3.98 & 4.21 & 0.24 & 4.29 \\
14 &11.44 & 9.40 & 4.01 & 0.01 & 4.77 \\
15 &12.31 & 10.05 & 4.05 & 0.05 & 3.63 \\
16 &12.60 & 51.52 & 4.02 & 0.01 & 3.91 \\
17 &12.82 & 1.49 & 3.93 & $3 \times 10^{-4}$ & 8.93 \\
18 &13.47 & 0.08 & 3.94 & $2 \times 10^{-3}$ & 4.25 \\
19 &13.63 & 0.14 & 4.14 & 0.01 & 4.16 \\
20 &14.37 & 0.27 & 3.98 & 0.02 & 3.92 \\
21 &15.01 & $5 \times 10^{-3}$ & 4.10 & $2 \times 10^{-5}$ & 3.20 \\
22 &16.71 & 0.02 & 4.87 & $7 \times 10^{-4}$ & 6.13 \\
23 &17.36 & 0.10 & 4.21 & 0.02 & 4.58 \\
\end{tabular}
\end{ruledtabular}
\end{table}
\end{center}
State 16 corresponds to excitation of the GT resonance that is reflected by 
the large $(p,n)$ cross section at zero-degree scattering being  51 mb/sr. The 
corresponding GT strength has a dimensionless value of 12.8. The states 7, 9, 
11, 15 and 16 have differential $(p,n)$ cross sections at zero-degree 
scattering larger than 10 mb/sr.  The ratio to the GT strength for these five 
states  are 4.16, 3.96, 3.47, 4.05 and 4.02. The ratio for these five leading 
states therefore lies between 3.47 and 4.16; and so there is about 20\% 
variation relative to the mean value. The six largest values for the 
$^{76}$Se$(n,p)$ cross section at zero-degree scattering are obtained for the 
states identified as  4, 7, 9, 10, 12 and 13 in the sequence. The ratios of 
these cross sections to their corresponding GT strengths vary between 3.1 and 
6.9.  Thus for these six largest $(n,p)$ transitions, there is a variation in 
the ratio of the zero-degree charge-exchange cross section to the GT strength 
of about 80\%. This is large in comparison to the variation in the ratios 
involving the strongest $(p,n)$ reaction cross sections. 

\begin{center}
\begin{figure}[ht]
\scalebox{0.45}{\includegraphics*{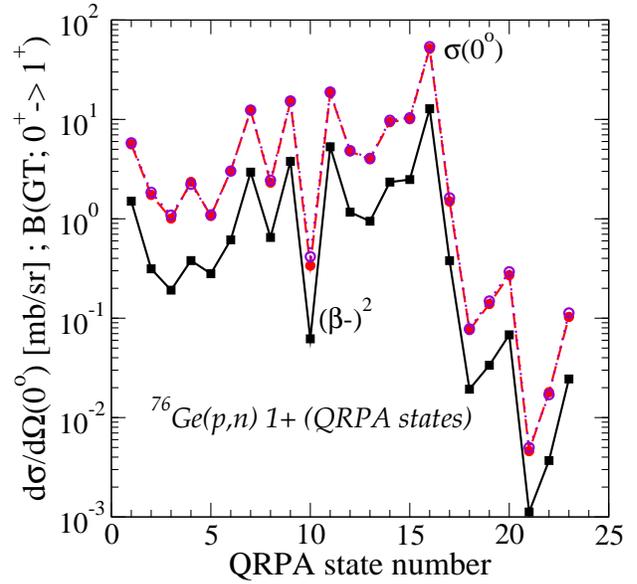}}
\caption{\label{FIG4}(color online)
differential cross sections at zero-degree scattering from DWA calculations of
$^{76}$Ge$(p,n){}^{76}$As$(1^+, m)$. Results found with and without non-central
components in the transition operator are depicted by the filled and open 
circles respectively. The filled squares are the values of 
$B(GT^{(-)})$ for $^{76}$Ge$\to ^{76}$As$(1^+, m)$ for each $1^+$ state.  }
\end{figure}
\end{center}

In fig.~\ref{FIG4} the zero-degree cross sections for $^{76}$Ge$(p,n)^{76}$As
$(1^+, m)$; $m = 1,\dots ,23$ are shown by the filled circles connected by 
dashed lines. The filled squares connected by solid lines are the dimensionless
GT strengths of the operator of eq.~(\ref{eq12}), $B(GT^{(-)},\, ^{76}$Ge
$\to ^{76}$As), for transition to each state of the QRPA given in sequence in 
table~\ref{200mev}.  The charge-exchange cross section values, connected by 
the dashed lines, resulted from DWA calculations made using the full $NN$ 
interaction (Melbourne force) as the transition operator. The open circles are
results obtained when only the central part of that transition operator was 
used. Clearly, for zero-degree scattering, the two-body spin-orbit and tensor 
contributions do not effect the cross sections appreciably.  The proton 
incident energy for all $(p,n)$ reactions is 200MeV.
Clearly the zero-degree cross sections for all $(p,n)$ transitions track 
similarly to the GT strengths of the same states. For the $(p,n)$, and 
presumably also for the corresponding reactions ($^3$He, $t$), the 
proportionality between those charge-exchange reaction cross sections in the 
forward direction and the GT strengths is fulfilled quite well. As noted above,
that means a proportionality within about 20\% for the five strongest 
transitions. 

\begin{center}
\begin{figure}[ht]
\scalebox{0.45}{\includegraphics*{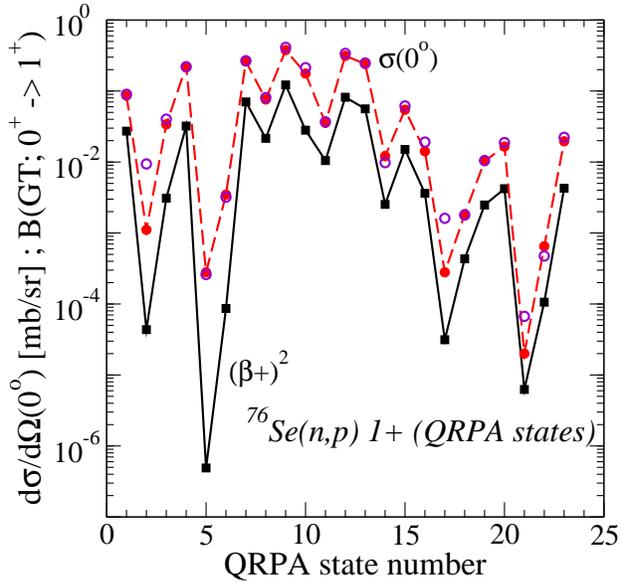}}
\caption{\label{FIG5}(color online)
differential cross sections at zero-degree scattering for $^{76}$Se$(n,p)$ to 
the 23 $1^+$ states in $^{76}$As (filled circles connected by dashed lines).
the DWBA results found by restricting the transition operator to just the 
central terms are depicted by the open circles. The associated GT strengths 
are shown by the filled squares connected by solid lines.}
\end{figure}
\end{center}

The situation is different for the $(n,p)$ reactions, and presumably also for 
the corresponding reactions $(t,^3$He) and $(^7$Li,$^7$Be). That is evident 
both from inspection of the results in table~\ref{200mev} and in 
fig.~\ref{FIG5}. For these transitions, most components are Pauli forbidden so
far as the GT operator is concerned.  Finite values occur only due to a 
smearing  of the Fermi surfaces. But the $(n,p)$ reactions, while hindered 
similarly, also can proceed by excitation of other components in the wave
functions. 
From fig.~\ref{FIG5} it is evident that the $(n,p)$ cross sections variation 
over the 23 QRPA possible states still tracks the values of the associated GT 
strengths.  However, note that both cross sections and GT strengths are much 
smaller than their counterparts in fig.~\ref{FIG4}, and the omission of 
non-central force elements in DWA calculations makes some greater variation 
than seen with the $(p,n)$ results.  Consequently, the proportionality 
`constant' of the ratio of forward direction, $(n,p)$ charge-exchange cross 
sections to the GT strengths for the six strongest transition is very large. 

\begin{center}
\begin{figure}[ht]
\scalebox{0.45}{\includegraphics*{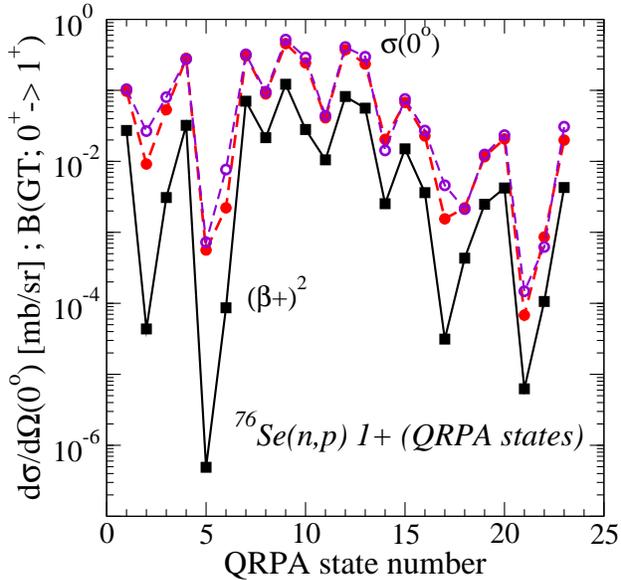}}
\caption{\label{FIG6}(color online)
zero-degree differential cross sections from DWA calculations of 120MeV 
$^{76}$Se$(n,p)$ to $1^+$ excited states in $^{76}$As and the associated 
GT strengths. Details are given in the text.}
\end{figure}
\end{center}

As the specifics of the Melbourne force change with incident energy, and
particularly those of the non-central components, we have made DWA calculations
at other energies. At 120MeV, the zero-degree cross section values  and their
ratios with the associated GT strengths are listed in table~\ref{120mev}.
As with the 200MeV results, the proportionality between the 
charge-exchange cross sections calculated at zero-degree scattering and the GT
strength is fulfilled to within about 20\% for the reaction $^{76}$Ge$(p,n)$.
However,  the variation is much larger for the ratios with the cross sections 
for $^{76}$Se$(n,p)$. In fact the variation of the latter ratios is near
90\% when one considers only the strongest $(n,p)$ transitions.
\begin{center}
\begin{table}[ht]
\begin{ruledtabular}
\caption{\label{120mev} 
the zero-degree charge-exchange cross sections and ratios (${\cal r}$) of 
them to the associated GT strength for the $^{76}$Ge$(p,n)$ $^{76}$As and 
$^{76}$Se$(n,p)$ $^{76}$As reactions to $^{76}$As$(1^+, m)$.  In this case, 
the incident energy was 120MeV.}
\vspace{0.2cm}
\begin{tabular}{|c|cc|cc|}
$m$ & ${}^{76}$Ge$(p,n) (0^\circ)$ & ${\cal r}$ & ${}^{76}$Se$(n,p) (0^\circ)$ 
& ${\cal r}$ \\
\hline
    1 &   6.51 &   4.33 &   0.10 &   3.58\\
    2 &   2.40 &   7.64 &  9x$10^{-3}$ & 210.10\\
    3 &   1.47 &   7.67 &   0.05 &  17.26\\
    4 &   2.98 &   7.85 &   0.28 &   8.80\\
    5 &   1.32 &   4.70 &  6x$10^{-4}$ &1146.9\\
    6 &   3.73 &   6.07 &  2x$10^{-3}$ &  25.56\\
    7 &  14.85 &   5.03 &   0.31 &   4.35\\
    8 &   2.77 &   4.26 &   0.09 &   4.13\\
    9 &  17.47 &   4.61 &   0.46 &   3.73\\
   10 &   0.46 &   7.45 &   0.24 &   8.68\\
   11 &  21.52 &   4.06 &   0.04 &   3.91\\
   12 &   5.65 &   4.84 &   0.37 &   4.56\\
   13 &   4.81 &   5.08 &   0.24 &   4.19\\
   14 &  11.07 &   4.73 &   0.02 &   8.06\\
   15 &  11.83 &   4.76 &   0.07 &   4.50\\
   16 &  61.63 &   4.81 &   0.02 &   6.29\\
   17 &   1.84 &   4.86 & 2x$10^{-3}$ &  49.43\\
   18 &   0.09 &   4.51 & 2x$10^{-3}$ &   5.12\\
   19 &   0.17 &   5.06 &   0.01 &   4.68\\
   20 &   0.33 &   4.87 &   0.02 &   4.91\\
   21 &  6x${10}^{-3}$ &  4.90 & 7x$10^{-5}$  &   10.88\\
   22 &   0.02 &   5.68 & 9x$10^{-4}$ &   8.10\\
   23 &   0.11 &   4.54 &   0.02 &   4.65\\
\end{tabular}
\end{ruledtabular}
\end{table}
\end{center}
In fig.~\ref{FIG6}, the zero-degree cross sections  from our DWA evaluations of
all $^{76}$Se$(n,p)$ reactions to the QRPA $1^+$ states in $^{76}$As are
shown for an incident neutron energy of 120MeV. As before, those results are
depicted by the filled and open circles (connected by the dashed lines to guide
the eye), with the filled circle presenting the results when the complete 
Melbourne force is used and the open circles giving the results when only the 
central force components are considered.  The associated GT strengths are 
depicted by the filled squares connected by the solid lines. As with the 
results for 200MeV, the $(n,p)$ cross section values vary across the 23 QRPA 
cases very similarly to the GT strength values.  But the devil is in the 
differences again and the ratio of them is far removed from being constant 
over the set.

\begin{center}
\begin{figure}[ht]
\scalebox{0.45}{\includegraphics*{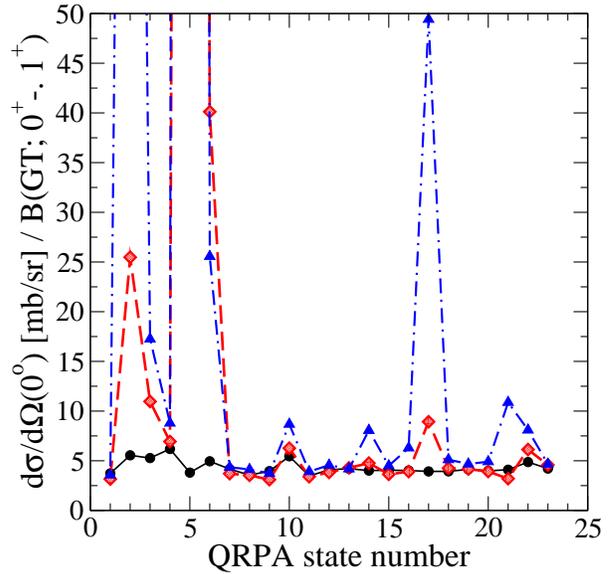}}
\caption{\label{FIG7}(Color online)
Ratios of cross sections from $^{76}$Ge$(p,n)$ $^{76}$As evaluated at zero 
degrees with the GT strengths (filled circles) for an incident 
energy of 200 MeV and for $^{76}$Se$(n,p)$ $^{76}$As displayed for 200 MeV 
(open diamonds) and for 120 MeV (filled triangles).}
\end{figure}
\end{center}

Finally, in figure~\ref{FIG7} the ratios of our calculated  zero-degree 
charge-exchange cross sections and GT strengths for the transition $^{76}$Ge 
$\to ^{76}$As (filled circles connected by a solid lines) to the 23 QRPA 
states in $^{76}$As and for the incident proton energy of 200MeV, are shown.
The ratios for $^{76}$Se$(n,p)$ $^{76}$As cross sections are displayed by the 
open diamonds connected by dashed lines (200MeV) and by the filled triangles 
connected by dot-dashed lines (120MeV). On this scale the relative smoothness
of the ratios for all 23 QRPA cases of $^{76}$Ge $\to ^{76}$As is apparent. 
The variation though is $\sim 20\%$.  But the extreme variation over the set 
for the $^{76}$Se$(n,p)$ $^{76}$As ratios makes it impossible to consider such
as forming a proportionality constant.  Even restricting consideration to the 
five strongest $(n,p)$ transitions yields a variation of $\sim 90 \%$.

\section{Conclusions}
\label{conclusions}

In conclusion, from our analysis of excitations with the QRPA model
structure for the three mass-76 nuclei, the 
proportionality between forward differential cross sections and GT strengths 
is fulfilled to within about 20\% for the ${}^{76}$Ge$(p,n)$ GT transitions. 
But this is not the case with forward differential 
charge-exchange $(n,p)$ reactions unlike the cases in light mass nuclei.
In light nuclei, where protons and
neutrons fill the same or nearly the same single-particle levels, the 
proportionality of forward charge-exchange reaction cross sections and the GT 
strengths seems valid for both $(n,p)$  and $(p,n)$ processes. For them the 
proton and neutron Fermi surfaces are similar.  But that is not so for medium 
heavy nuclei such as the mass-76 set we considered herein. 
With them the proton and neutron Fermi surfaces are quite 
different, and the effects of Pauli-blocking (hindering for smeared 
surfaces) allows the proportionality to be good (within 20\%) for $(p,n)$
processes but not for the $(n,p)$ transitions. This arises
because the charge-exchange $(n,p)$ transitions are sensitive to radial
overlaps of single particle wave functions whereas the GT values are
not, and for the $(n,p)$ processes those in which a nucleon stays within 
a given orbit but changes type are not as dominant as with the $(p,n)$
ones.

We anticipate that such will be the case for systems that have sizeable 
neutron excess. It will be interesting not only to apply the approach we
have taken in cases whenever relevant data are available with which 
additional tests of the quality of the nuclear structure model can be made but
also in any case when application of theories for
charge-exchange processes initiated by composite projectiles, as
comparably physically justified, can be made. 


\begin{acknowledgments}
We are most grateful to  Prof. Herbert M\"uther for providing us with 
solutions of the Bethe-Goldstone equation starting with the Bonn-CD $NN$ 
force.  The work of V. R. has been supported by the Deutsche 
Forschungsgemeinschaft (by grant FA67/28-2 and within the Transregio Project 
TR27 ``Neutrinos and Beyond").  A. F. and V. R. thank also the EU ILIAS 
project under the contract RII3-CT-2004-506222 for  support.
\end{acknowledgments}

\bibliography{2side_ver3}

\end{document}